\documentclass[twocolumn,superscriptaddress]{revtex4}
\usepackage{graphicx}
\usepackage{color}

\begin{document}

\title{Efficient vortex generation in sub-wavelength epsilon-near-zero slabs}

\author{Alessandro Ciattoni}
\affiliation{Consiglio Nazionale delle Ricerche, CNR-SPIN, Via Vetoio 10, 67100 L'Aquila, Italy}

\author{Andrea Marini}
\affiliation{ICFO-Institut de Ciencies Fotoniques, The Barcelona Institute of Science and Technology, 08860 Castelldefels (Barcelona), Spain}

\author{Carlo Rizza}
\affiliation{Consiglio Nazionale delle Ricerche, CNR-SPIN, Via Vetoio 10, 67100 L'Aquila, Italy}

\begin{abstract}
We show that a homogeneous and isotropic slab, illuminated by a circularly polarized beam with no topological charge, produces vortices of order two in the opposite circularly polarized components of the reflected and transmitted fields, as a consequence of the difference between transverse magnetic and transverse electric dynamics. In the epsilon-near-zero regime, we find that vortex generation is remarkably efficient in sub-wavelength thick slabs up to the paraxial regime. This physically stems from the fact that a vacuum paraxial field can excite a nonparaxial field inside an epsilon-near-zero slab since it hosts slowly varying fields over physically large portion of the bulk. Our theoretical predictions indicate that epsilon-near-zero media hold great potential as nanophotonic elements for manipulating the angular momentum of the radiation, since they are available without resorting to complicated micro/nano fabrication processes and can operate even at very small (ultraviolet) wavelengths.
\end{abstract}

\maketitle

Spin-orbit interaction (SOI) of light is a very important research topic since it provides a tool for manipulating the spatial degrees of freedom of the radiation by acting on its circular polarization state \cite{Bliok1}. A remarkable SOI effect is the generation of optical vortices from circularly polarized beams, a process accompanied by spin to orbital angular momentum conversion. Standard procedures to achieve vortex generation are focusing by high-numerical aperture lens \cite{Zhaooo,Bliok2}, scattering by small particles \cite{Bliok2}, propagation along the optical axis of a uniaxial crystal \cite{Ciatt1,Brass1} and propagation through semiconductor microcavities \cite{Mannii}. Similar SOI effects involving Bessel beams have been considered in uniaxial crystals \cite{Khilo1} and at reflection and transmission by a planar interface between two homogeneous media \cite{Yavors}. The advent of metamaterials has further increased the SOI research effort \cite{Shitri}, mostly in the use of ultra-thin metasurfaces for manipulating the angular momentum of light \cite{Liiiii,Hakoby} and for vortex generation \cite{Yanggg,Chennn}.

Epsilon near zero (ENZ) media are nowadays attracting an increasing research interest due to the very unconventional way they affect the electromagnetic radiation. The effective wavelength in ENZ media is much larger than the vacuum wavelength and this entails a regime quite opposite to geometrical optics where the field is slowly-varying over relatively large portions of the bulk. Such feature has been exploited for squeezing electromagnetic waves at will \cite{Silve1}, for tailoring the antenna radiation pattern \cite{Aluuu1} and for enhancing nonlinear response of matter \cite{Ciatt2,Argyr1,Ciatt3,Alammm,Ciatt4}. In the context of light SOI, it has recently been proposed that a thin epsilon-near-zero slab can enhance the spin Hall effect of transmitted light \cite{Zhuuuu}.

In this letter we show that a homogeneous, isotropic and ultra-thin (sub-wavelength thick) slab can support vortex generation. We prove that such genuine SOI effect is physically due to the mutual difference between the dynamics of transverse magnetic and transverse electric fields upon reflection and transmission. As the majority of radiation SOI phenomena, the slab vortex generation is mainly a nonparaxial effect. On the other hand, we prove that slab vortex generation in the ENZ regime is remarkably efficient even for incident paraxial beams in spite of the very small slab thickness. Such phenomenology is unprecedent since, to the best of our knowledge, paraxial vortex generation in homogenous media (i.e. through lenses and uniaxial crystals) requires samples whose thickness is much larger than wavelength. Here, the crucial role is played by the physical ability of an ENZ slab to turn a paraxial wave, incoming from vacuum, into a nonparaxial one within the bulk, its nonparaxiality triggering the predicted slab vortex generation. The vortex generation method prosed in this letter can have important nanophotonic applications since, unlike the one based on metasurfaces, it does not require microfabrication (the ENZ slab is homogenous) and it is scalable down to very small wavelengths (exploiting the ultraviolet ENZ point of metals) where metamaterials are not available.

Let us consider the scattering of a monochromatic ($\sim \exp(-i \omega t )$) electromagnetic field by an homogeneous and isotropic slab of thickness $L$ and dielectric permittivity $\varepsilon$ (see Fig.1). We choose the $z$ axis to be along the slab normal. The angular spectrum representations of both the incident $(i)$ and transmitted $(t)$ fields are ($q=i,t$)
\begin{eqnarray} \label{ang}
{\bf{E}}^{(q)} &=& \int {d^2 {\bf{k}}_ \bot  } e^{ i{\bf{k}}_\bot   \cdot {\bf{r}}_ \bot +i k_z^{(V)} z } \nonumber \\
&\times& \left[ \left( \frac{{k_x {\bf{\hat e}}_x  + k_y {\bf{\hat e}}_y }}{{k_ \bot  }} - \frac{ k_\bot {\bf{\hat e}}_z  }{{k_z^{(V)} }}  \right) U_{TM}^{(q)} \right. \nonumber \\
  &+& \left. \left( {\frac{{ - k_y {\bf{\hat e}}_x  + k_x {\bf{\hat e}}_y }}{{k_ \bot  }}} \right) U_{TE}^{(q)}  \right],
\end{eqnarray}
where ${\bf{r}}_ \bot   = x{\bf{\hat e}}_x  + y{\bf{\hat e}}_y$ and ${\bf{k}}_ \bot   = k_x{\bf{\hat e}}_x  + k_y{\bf{\hat e}}_y$ are the transverse position and wavevector, respectively, $k_ \bot   = \sqrt{k_x^2 + k_y^2}$, $k_z^{\left( V \right)} \left( {k_ \bot  } \right) = \sqrt {k_0^2  - k_ \bot ^2 }$ is the longitudinal vacuum wave vector ($k_0 = \omega / c$) and $U_{TM}^{(q)} \left( {{\bf{k}}_ \bot  } \right)$ and $U_{TE}^{(q)} \left( {{\bf{k}}_ \bot  } \right)$ are the amplitudes of the transverse magnetic (TM) and transverse electric (TE) components. The slab differently transmits TM and TE fields according to $U_{TM}^{\left( t \right)}  = t_{TM} \left( k_\bot \right) U_{TM}^{\left( i \right)}$ and $U_{TE}^{\left( t \right)}  = t_{TE} \left( k_\bot \right) U_{TE}^{\left( i \right)}$ where the complex transmissivities are
\begin{eqnarray} \label{tt}
 t_{TM}  &=& \left[ {\cos \left( {k_z^{\left( S \right)} L} \right) - \frac{i}{2}\left( {\frac{{\varepsilon k_z^{\left( V \right)} }}{{k_z^{\left( S \right)} }} + \frac{{k_z^{\left( S \right)} }}{{\varepsilon k_z^{\left( V \right)} }}} \right)\sin \left( {k_z^{\left( S \right)} L} \right)} \right]^{ - 1}, \nonumber \\
 t_{TE}  &=& \left[ {\cos \left( {k_z^{\left( S \right)} L} \right) - \frac{i}{2} \left({ \frac{{k_z^{\left( V \right)} }}{{k_z^{\left( S \right)} }} + \frac{{k_z^{\left( S \right)} }}{{k_z^{\left( V \right)} }}} \right)\sin \left( {k_z^{\left( S \right)} L} \right)} \right]^{ - 1}, \nonumber \\
\end{eqnarray}
where $k_z^{\left( S \right)} \left( {k_ \bot  } \right) = \sqrt {k_0^2 \varepsilon - k_ \bot ^2 }$ is the longitudinal wave vector inside the slab. We now choose, as a basis for the transverse plane, the left handed circular (LHC) and right handed circular (RHC) polarization unit vectors, ${\bf{\hat e}}_L  = \frac{1}{\sqrt{2}} \left( {{\bf{\hat e}}_x  + i{\bf{\hat e}}_y } \right)$ and ${\bf{\hat e}}_R  = \frac{1}{\sqrt{2}} \left( {{\bf{\hat e}}_x  - i{\bf{\hat e}}_y } \right)$, respectively, and we introduce polar coordinates, $x = r_ \bot  \cos \varphi$, $y = r_ \bot  \sin \varphi$, $k_x  = k_ \bot  \cos \theta$, $k_y  = k_ \bot  \sin \theta$, for both transverse position and wave vector. Accordingly Eq.(\ref{ang}) becomes
\begin{eqnarray} \label{ang_pol}
{\bf{E}}^{\left( q \right)}  &=& \int\limits_0^{ + \infty } {dk_ \bot  } k_ \bot  \int\limits_0^{2\pi } {d\theta } \;
e^ { ik_ \bot  r_ \bot  \cos \left( {\theta  - \varphi } \right) +ik_z^{\left( V \right)} z } \nonumber \\
&\times& \left[ e^ { -i\theta } \left( {\frac{{U_{TM}^{\left( q \right)}  - iU_{TE}^{\left( q \right)} }}{{\sqrt 2 }}} \right){\bf{\hat e}}_L  \right. \nonumber \\
&+& \left.  e^ { i\theta } \left( {\frac{{U_{TM}^{\left( q \right)}  + iU_{TE}^{\left( q \right)} }}{{\sqrt 2 }}} \right){\bf{\hat e}}_R + \left( { - \frac{{k_ \bot  U_{TM}^{\left( q \right)}}}{{k_z^{\left( V \right)} }} } \right){\bf{\hat e}}_z  \right]. \nonumber \\
\end{eqnarray}
\begin{figure}
\center
\includegraphics[width=0.48\textwidth]{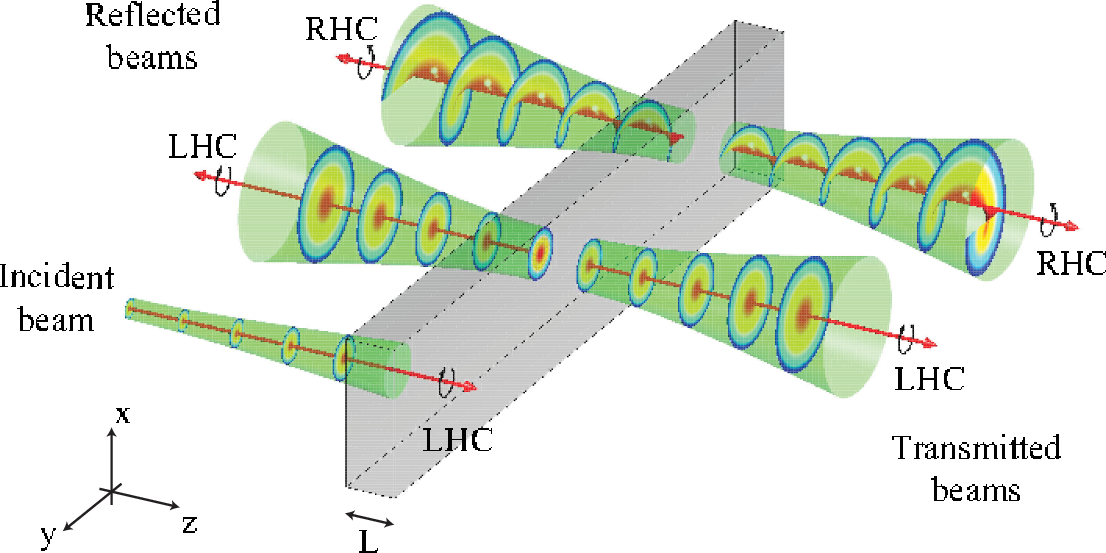}
\caption{(Color online). Geometry of the vortex generation process. The incident beam is a LHC polarized beam with no topological charge. Both the reflected and the transmitted fields have LHC and RHC polarized components, the former being circularly symmetric and the latter containing a vortex of the second-order (The spatial separation in figure among such components is introduced for clarity purposes).}
\end{figure}
The angular factors $e^{-i \theta} = \frac{1}{k_\bot } \left( k_x -i k_y  \right)$ and $e^{i \theta }= \frac{1}{k_\bot } \left( k_x + i k_y  \right)$ of Eq.(\ref{ang_pol}) arise from the TM and TE polarizations and hence they encode the field transversality $\nabla \cdot {\bf E} = 0$ in the chosen circular basis. To investigate the slab spin to orbital conversion of radiation angular momentum, we consider an incident beam which is LHC polarized (carrying spin angular momentum) and which has no topological charge (i.e. not carrying orbital angular momentum). From Eq.(\ref{ang_pol}), it follows that its TM and TE spectral amplitudes have to satisfy the constraint $U_{TE}^{\left( i \right)} = i U_{TM}^{\left( i \right)}$ which, together with the zero-topological charge condition, implies that $U_{TM}^{\left( i \right)} =  e^{i\theta } U^{\left( i \right)}\left( {k_ \bot  } \right)$ and $U_{TE}^{\left( i \right)} = i e^{i\theta } U^{\left( i \right)} \left( {k_ \bot  } \right)$, where $U^{\left( i \right)}$ is an arbitrary rotationally invariant spectrum. As a matter of fact, Eq.(\ref{ang_pol}) with $q=i$, after performing the angular integration, yields
\begin{eqnarray} \label{in}
{\bf{E}}^{\left( i \right)}  &=& 2\pi \int\limits_0^{ + \infty } {dk_ \bot  } k_ \bot  e^{ ik_z^{\left( V \right)} z } \left[ \sqrt{2} J_0 \left( k_ \bot  r_ \bot   \right) {\bf{\hat e}}_L  \right. \nonumber \\
&+& \left.  e^{i\varphi }  \left( {\frac{{k_ \bot  }}{{ik_z^{\left( V \right)} }}} \right) J_1 \left( {k_ \bot  r_ \bot  } \right) {\bf{\hat e}}_z \right] U^{\left( i \right)}.
\end{eqnarray}
where $J_n (\xi )$ is the Bessel function of the first kind of order $n$. The TM and TE components of the field transmitted by the slab are therefore $U_{TM}^{\left( t \right)}  = t_{TM} e^{i\theta } U^{\left( i \right)}$ and $U_{TE}^{\left( t \right)}  = t_{TE} ie^{i\theta } U^{\left( i \right)}$ and accordingly Eq.(\ref{ang_pol}) with $q=t$, after performing the angular integration, yields
\begin{eqnarray} \label{tr}
{\bf{E}}^{\left( t \right)}  &=& 2\pi \int\limits_0^{ + \infty } {dk_ \bot  } k_ \bot  e^{ ik_z^{\left( V \right)} z } \left[ \left( {\frac{{t_{TM}  + t_{TE} }}{{\sqrt 2 }}} \right)J_0 \left( {k_ \bot  r_ \bot  } \right){\bf{\hat e}}_L \right. \nonumber \\
 &-& \left. e^{ i2\varphi} \left( {\frac{{t_{TM}  - t_{TE} }}{{\sqrt 2 }}} \right)J_2 \left( {k_ \bot  r_ \bot  } \right){\bf{\hat e}}_R \right. \nonumber \\
 &+& \left. e^{ i\varphi } \left( {\frac{{k_ \bot t_{TM} }}{{ik_z^{\left( V \right)} }} } \right)J_1 \left( {k_ \bot  r_ \bot  } \right){\bf{\hat e}}_z  \right]U^{\left( i \right)}.
\end{eqnarray}
Equation (\ref{tr}) reveals that the RHC component of the transmitted field, due to the factor $\exp (i2\varphi)$, contains a second-order vortex (i.e its topological charge is equal to $2$) whereas the LCH component has no topological charge as the incident field. Evidently the reflected beam has the same polarization and vortex structure. The geometry of the incident, reflected and transmitted fields is sketched in Fig.1. In addition, Eq. (\ref{tr}) clearly shows that the generation of the transmitted RHC field (containing the vortex) is a consequence of condition $t_{TM} \neq t_{TE}$, i.e. of the different behavior of TM and TE fields upon slab transmission. Physically this is due to the fact that the TM and TE spectral amplitudes of the incident LHC polarized field are related by the constrain $U_{TE}^{\left( i \right)} = i U_{TM}^{\left( i \right)}$ which, due to the different slab effect on TM and TE fields, is not transferred to the transmitted field which accordingly in not LHC polarized. A fundamental role in the considered vortex generation process is played by the factors $e^{-i \theta}$ and $e^{i \theta}$ of Eq.(\ref{ang_pol}). As a matter of fact the topological charge factor $e^{i 2 \varphi}$ appears (after the integration on $\theta$) since the $e^{i \theta}$ factor is literally carried from the spectral amplitude of the incident field $e^{i \theta} U^{(i)}$ to the RHC component of the transmitted field where it is multiplied by the further polarization factor $e^{i \theta}$. It is worth noting that an analogous vortex generation mechanism, also based on the above TM-TE interplay, occurs in spatially homogeneous unbounded uniaxial crystal where the vortex is produced by the propagation dephasing between extraordinary and ordinary components which (due to the rotational invariance around the optical axis) are TM and TE waves, respectively \cite{Ciatt1}. Remarkably, even the spin-to-orbital angular momentum conversion in semiconductor microcavities is due to a kind of TM-TE polarization splitting \cite{Mannii}.
\begin{figure}
\center
\includegraphics[width=0.48\textwidth]{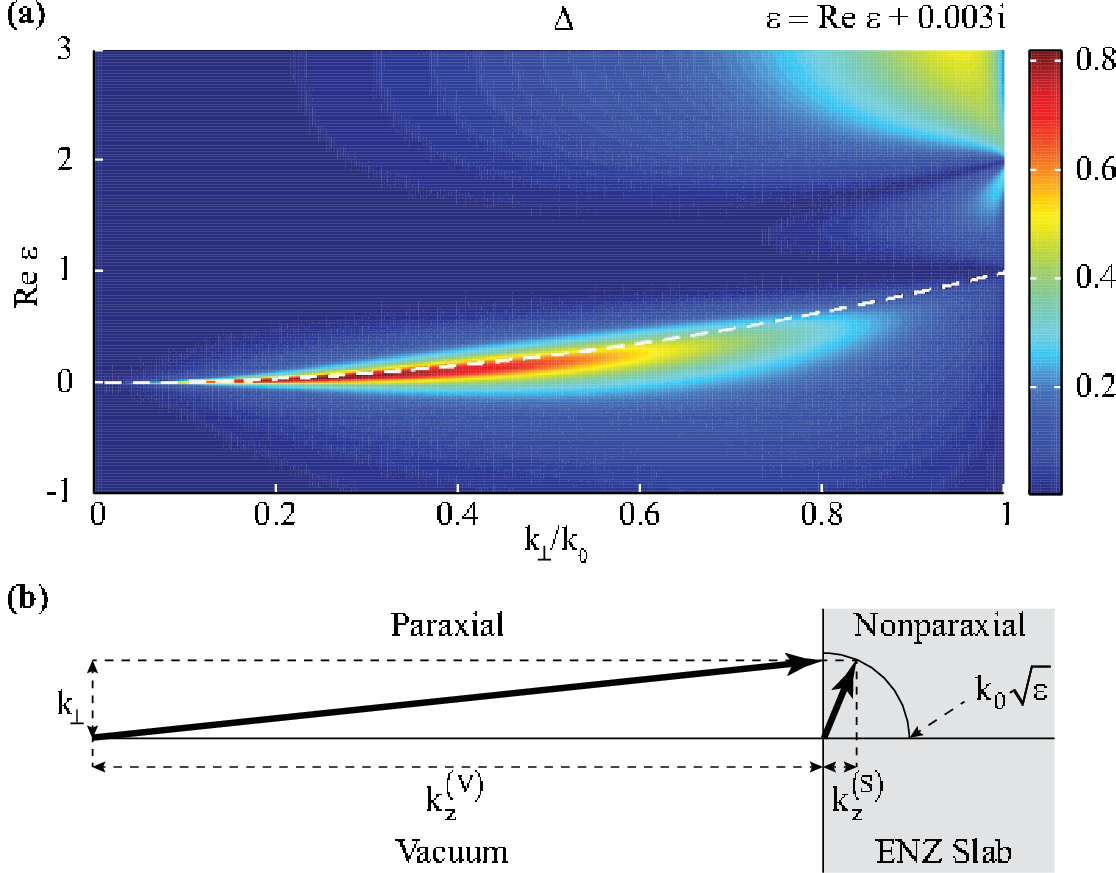}
\caption{(Color online) (a) Vortex spectral amplitude $\Delta = |t_{TM} - t_{TE}|$ versus the normalized transverse wave vector $k_\bot$ and the real part of the slab permittivity $\varepsilon = {\rm Re} \: \varepsilon  + 0.003i$. The slab thickness is half the vacuum wavelength. The white dashed line is the curve $k_\bot = k_0 \sqrt{ {\rm Re} \: \varepsilon}$ on which $\left| k_z^{\left( S \right)} \right| = k_0 \sqrt{ {\rm Im} \: \varepsilon}$ is minimum. (b) Wave vectors diagram showing the excitation of nonparaxial waves within the ENZ slab by vacuum paraxial waves (Here ${\rm Im} \: \varepsilon = 0$ for simplicity).}
\end{figure}

To investigate the vortex generation process in more detail, we consider slabs of permittivity $\varepsilon = {\rm Re} \: \varepsilon + 0.003i$ and sub-wavelength thickness $L = \frac{1}{2} \lambda$ ($\lambda = \frac{2\pi c}{\omega}$ is the vacuum wavelength) and we focus on the absolute value of the difference between the TM and TE transmissivities of Eqs.(\ref{tt}), $\Delta = |t_{TM} - t_{TE}|$. This is the key spectral parameter ruling vortex generation in the transmitted RHC component of Eq.(\ref{tr}). In Fig.2(a) we plot $\Delta$ as a function of the real part of the slab permittivity and of the transverse wave vector $k_\bot$ (normalized with the vacuum wave number $k_0$) spanning the vacuum homogeneous spectrum ($k_\bot < k_0$). For standard materials, ${\rm Re} \: \varepsilon > 1$, $\Delta$ is rather small except for $k_\bot$ close to $k_0$. This is consistent with the general nonparaxial trait of SOI optical effects \cite{Bliok1}. For materials characterized by the condition $|{\rm Re} \: \varepsilon | < 1$ the situation is remarkably different in that $\Delta$ has a marked lobe, localized at the middle of the vacuum homogeneous spectrum, whose left tail encompasses very small transverse wave vectors $k_\perp \ll k_0$ around ${\rm Re} \: \varepsilon = 0$. This proves that vortex generation in ENZ sub-wavelength thick slabs can efficiently be observed even in the paraxial regime, the smaller ${\rm Re} \: \varepsilon$ the more paraxial the field.
\begin{figure}
\center
\includegraphics[width=0.48\textwidth]{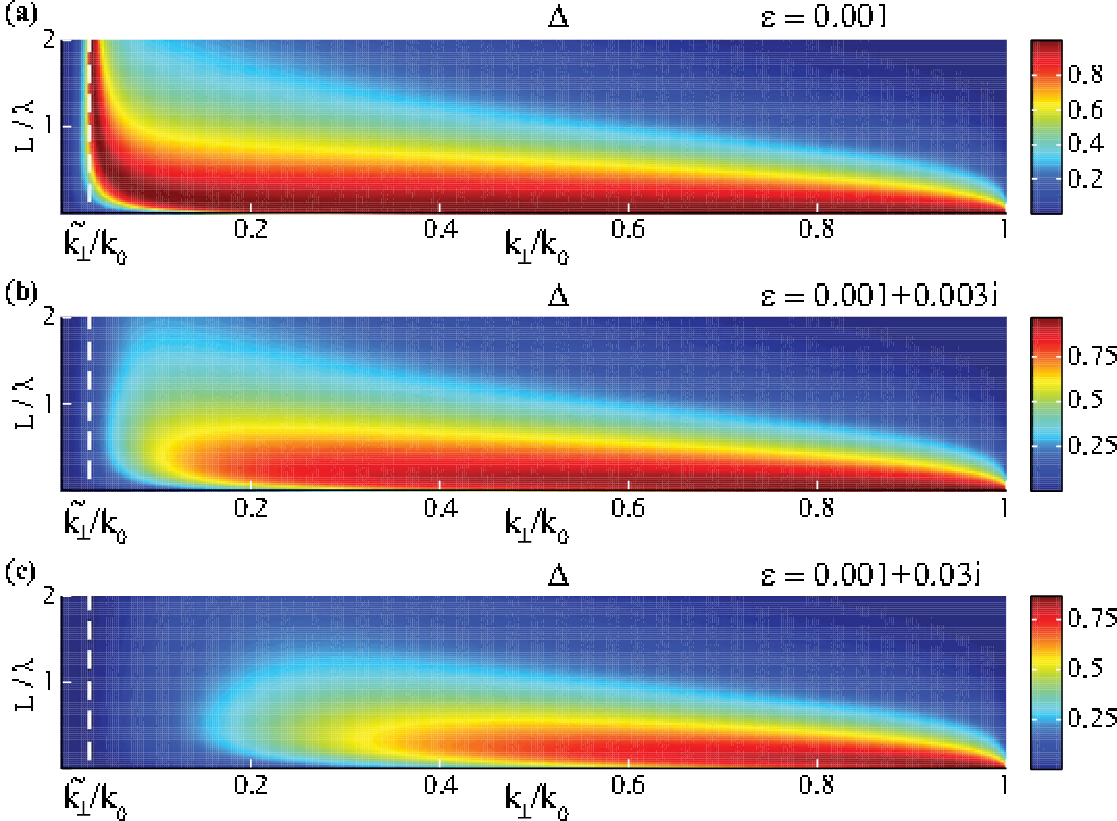}
\caption{(Color online) Vortex spectral amplitude $\Delta$ versus the normalized transverse wave vector $k_\bot$ and the normalized slab thickness $L$, for three different slabs with permittivities (a) $\varepsilon = 0.001$, (b) $\varepsilon = 0.001+0.003i$ and (c) $\varepsilon = 0.001+0.03i$. The white dashed lines indicate the transverse wave vector $\tilde{k}_\bot = k_0 \sqrt{{\rm Re} \: \epsilon} \simeq 0.031 k_0$}
\end{figure}
To physically grasp the mechanism supporting paraxial vortex generation in ENZ slabs, in Fig.2(a) we have also plotted the curve $k_\bot = k_0 \sqrt{ {\rm Re} \: \varepsilon }$ (white dashed line) which is found to locate the largest values of $\Delta$ on the lobe. Since $\left| k_z^{\left( S \right)} \right| = k_0 \left[ \left( {\rm Re} \: \varepsilon -  \frac{k_\bot^2}{k_0^2}  \right)^2+ \left( {\rm Im} \: \varepsilon  \right)^2 \right]^{1/4}$, it is evident that the longitudinal wave vector within the slab attains its minimum absolute value $\left| k_z^{\left( S \right)} \right|_{min} = k_0 \sqrt{ {\rm Im} \: \varepsilon}$ at all points of the curve. Therefore vortex generation is efficient in the paraxial regime since, if $0 < {\rm Re} \: \varepsilon  < 1$, there is a narrow bundle of vacuum paraxial waves whose $k_\bot$ is smaller than and close to $k_0 \sqrt{ {\rm Re} \: \varepsilon }$ and they excite highly nonparaxial waves within the slab, their wave vector being almost orthogonal to the $z$-axis. In Fig.2(b) we have sketched a wave vectors diagram showing the excitation of nonparaxial waves in the ENZ slab by paraxial vacuum waves in the ideal ${\rm Im} \: \varepsilon  = 0$ situation. It is evident that the smaller the value $\left| k_z^{\left( S \right)} \right|_{min} = k_0 \sqrt{ {\rm Im} \: \varepsilon}$ the more nonparaxial are the waves within the slab so that the full ENZ condition $|\varepsilon| \ll 1$  has to be met to achieve efficient vortex generation. The imaginary part of $\varepsilon$, due to the slab losses, also play an important role as combined to the slab thickness $L$. In Fig.3 we plot the vortex spectral amplitude $\Delta$ as a function of the slab thickness $L$ (normalized with the vacuum wavelength $\lambda$) and of the transverse wave vector $k_\bot$ (normalized with the vacuum wave number $k_0$) spanning the vacuum homogeneous spectrum ($k_\bot < k_0$), for three different slabs with permittivities (a) $\varepsilon = 0.001$, (b) $\varepsilon = 0.001+0.003i$ and (c) $\varepsilon = 0.001+0.03i$. In the first (ideal) case paraxial vortex generation around $\tilde{k}_\bot = k_0 \sqrt{{\rm Re} \: \epsilon} \simeq 0.031 k_0$ (labelled with the vertical white dashed line) is persistent even for $L$ greater than $\lambda$. This is due to the fact that in this case, from Eqs.(\ref{tt}) we have $t_{TM}\left(\tilde{k}_\bot\right)  = \left( {1 - \frac{i}{2}\varepsilon \sqrt {1 - \varepsilon } k_0 L} \right)^{ - 1}$ and $t_{TE} \left(\tilde{k}_\bot\right) = \left( {1 - \frac{i}{2}\sqrt {1 - \varepsilon } k_0 L} \right)^{ - 1}$ so that, since $\varepsilon \ll 1$, the TM transmissivity is slowly varying with $L$ and very close to $1$ whereas the TE transmissivity quickly decreases as $L$ increases thus producing a large value of $\Delta$. In the other two cases (see Figs.3(b) and 3(c)), the imaginary part is larger and its detrimental effect on vortex generation is evident since $\Delta$ fades as the slab thickness $L$ increases. This is due to the fact that absorption in the ENZ regime is generally not negligible even if ${\rm Im} \varepsilon$ is rather small. However, if the slab has sub-wavelength thickness, vortex generation persists even up to the paraxial regime.

\begin{figure}
\center
\includegraphics[width=0.48\textwidth]{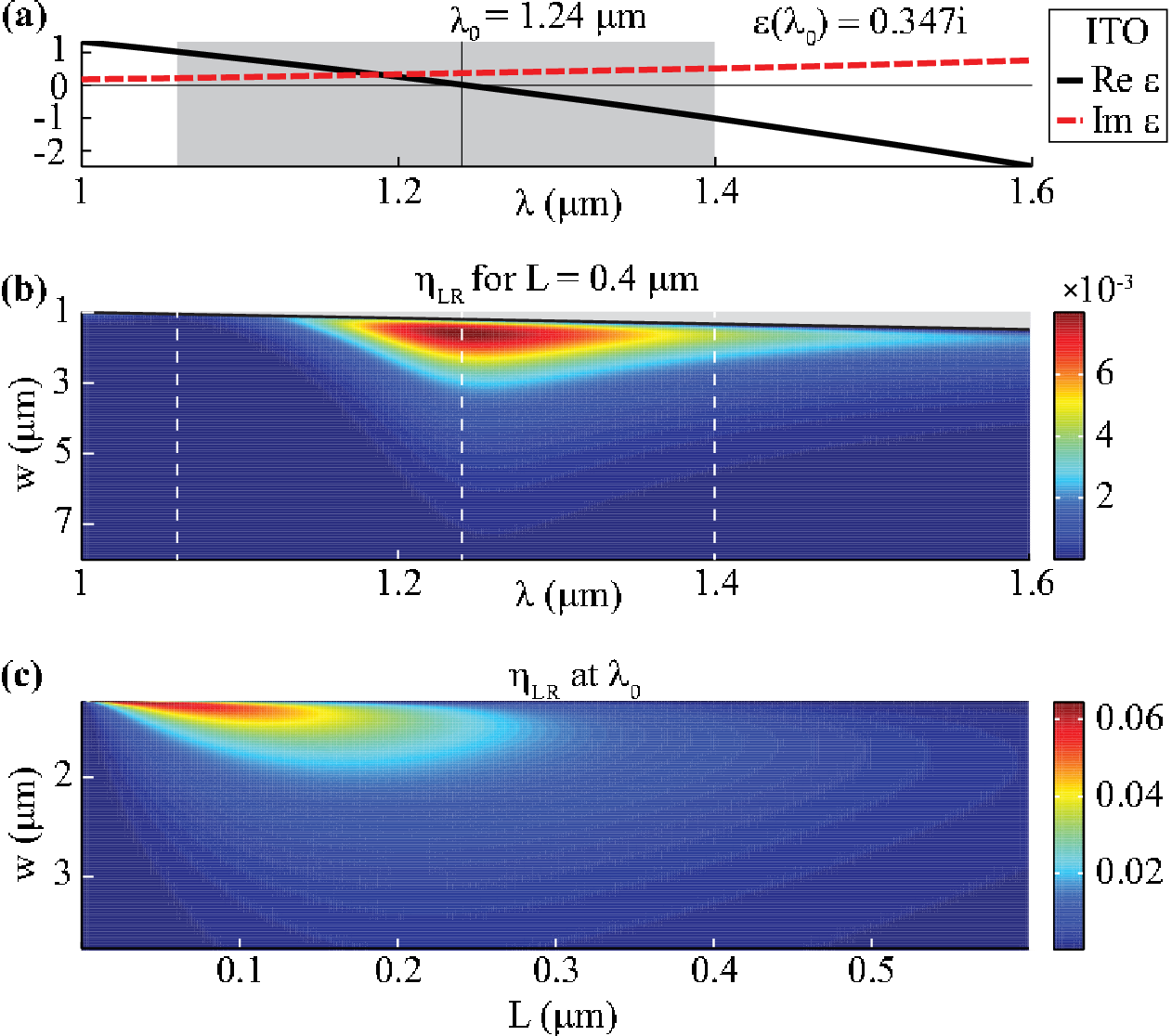}
\caption{(Color online) (a) Real and imaginary part of indium tin oxide (ITO) permittivity in the infrared spectral band $1 \: \mu {\rm m} < \lambda < 1.6 \: \mu {\rm m}$. The shadowed region corresponds to $| {\rm Re} \: \varepsilon | < 1$. (b) Vortex generation efficiency $\eta_{LR}$ for Bessel beams by an ITO slab of thickness $L=0.4 \: \mu{\rm m}$ as a function of the vacuum wavelength $\lambda = 2\pi c/\omega$ and the beam width $w$. In the upper triangular gray region, $w < \lambda$ and Bessel beams are evanescent. (c) Vortex generation efficiency $\eta_{LR}$ at $\lambda_0$ by ITO slabs as a function of the slab thickness $L$ and the beam width $w$.}
\end{figure}

In order to discuss the vortex generation process in realistic and feasible situations, we consider sub-wavelength thick slabs whose dielectric dispersive response is described by the Drude model $\varepsilon  = \varepsilon _\infty   - \frac{\omega _p^2} {\omega ^2  + i\gamma \omega }$ where $\varepsilon _\infty$ is the high-frequency permittivity, $\gamma$ is the damping rate and $\omega_p$ is the free-electron plasma frequency. Such materials have the zero crossing point ${\rm Re} \: \varepsilon = 0$ at vacuum wavelength $\lambda _0  = 2\pi c\sqrt{\varepsilon _\infty} \left( \omega _p^2  - \gamma ^2 \varepsilon _\infty \right)^{-1/2}$ where ${\mathop{\rm Im}\nolimits} \: \varepsilon  =  \gamma \varepsilon _\infty^{3/2} \left( \omega _p^2  - \gamma ^2 \varepsilon _\infty \right)^{-1/2}$. For simplicity we consider an incident LCH polarized Bessel beam whose LHC component is $E_L^{\left( i \right)}  = e^{i\sqrt {k_0^2  - \left( \frac{2\pi}{w} \right)^2 } z} J_0 \left( \frac{2\pi r_ \bot}{w} \right)A^{\left( i \right)}$ where $w= 2\pi/ \delta k$ is the beam width related to the transverse spectral semi-width $\delta k$ of the Bessel cone and $A^{(i)}$ is an arbitrary amplitude. Such field is obtained from Eq.(\ref{in}) for the spectral amplitude $U^{\left( i \right)} =  \frac{A^{\left( i \right)} w}{ 2^{5/2} \pi^2}  \delta \left( {k_ \bot   - \frac{2\pi}{ w}} \right)$ so that, from Eq.(\ref{tr}), the RHC component of the transmitted field is $E_R^{\left( t \right)}  = e^{i\sqrt {k_0^2  - \left( \frac{2\pi}{w} \right)^2 } z + i2\varphi } J_2 \left( \frac{2\pi r_ \bot}{w} \right) \frac{1}{2}\left[ t_{TE} \left( \frac{2\pi}{w} \right) - t_{TM} \left( \frac{2\pi}{w} \right) \right] A^{\left( i \right)}$. The efficiency $\eta_{LR}$ of the vortex generation process is measured by the portion of the incident power (carried by a LCH polarized beam) which flows into the produced vortex (carried by a RHC polarized beam). Accordingly we set $\eta_{LR} = \mathop{\lim }\limits_{\rho  \to  + \infty } \frac{P^{(t)}_R (\rho)}{P^{(i)}_L (\rho)}$ where $P^{(q)}_C (\rho)= \pi \int_0^\rho d r_\bot r_\bot \hat{\bf e}_z \cdot {\rm Re}\left( {\bf E}^{(q)}_C \times {{\bf H}^{(q)}_C}^* \right)$ is the power carried by the $C=L,R$ polarized component of the $q =i,t$ beam and flowing through a disk of radius $\rho$ coaxial with the $z$-axis \cite{Yavors}. For the above considered bessel beam scattering, after calculations, we get $\eta_{LR} =\frac{1}{4} |t_{TE} \left( \frac{2\pi}{w} \right) - t_{TM} \left( \frac{2\pi}{w} \right)|^2$.
\begin{figure}
\center
\includegraphics[width=0.48\textwidth]{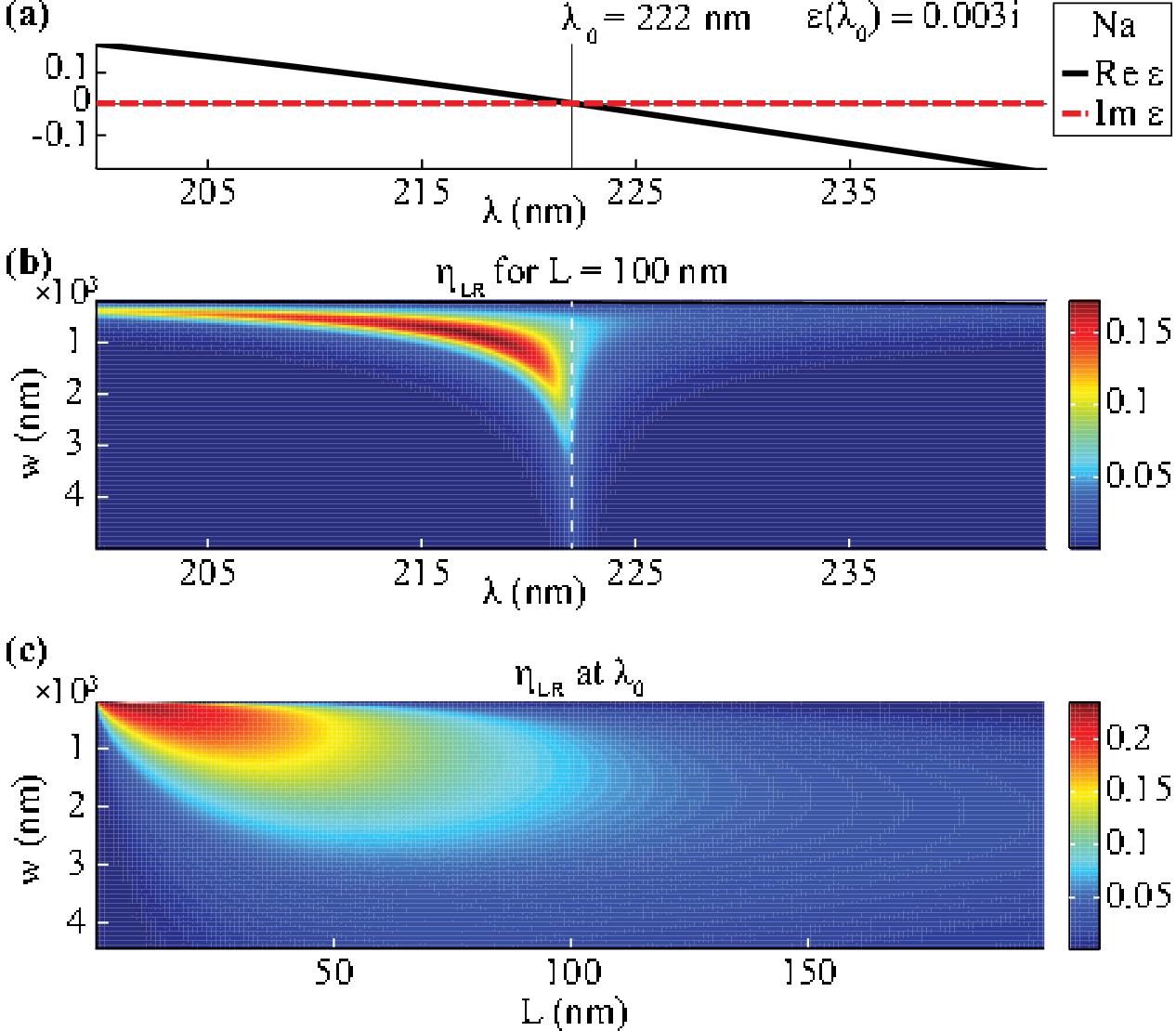}
\caption{(Color online) (a) Real and imaginary part of sodium (Na) permittivity in the ultraviolet spectral band $200 \: {\rm nm} < \lambda < 245 \: {\rm nm}$. (b) Vortex generation efficiency $\eta_{LR}$ for Bessel beams by a Na slab of thickness $L=100 \: {\rm nm}$ as a function of the vacuum wavelength $\lambda = 2\pi c/\omega$ and the beam width $w$. (c) Vortex generation efficiency $\eta_{LR}$ at $\lambda_0$ by Na slabs as a function of the slab thickness $L$ and the beam width $w$.}
\end{figure}

In the first example we consider an indium tin oxide (ITO) slab of thickness $L = 0.4 \: \mu {\rm m}$ whose Drude parameters are $\epsilon_\infty = 3.8055$, $\omega_p = 2.9719 \times 10^{15}$ Hz and $\gamma = 0.0468 \: \omega_p$ \cite{Alammm}, for which $\lambda_0 = 1.24 \: \mu {\rm m}$ and ${\rm Im} \left[ \varepsilon (\lambda_0) \right] = 0.347$. In Fig.4(a) we plot the real and imaginary parts of ITO permittivity in the infrared spectral band $1 \: \mu {\rm m} < \lambda < 1.6 \: \mu {\rm m}$ and we have shadowed the spectral region $| {\rm Re} \: \varepsilon | < 1$ where, from the above discussion, vortex generation is expected to occur even in the paraxial regime. In Fig.4(b) we plot the Bessel vortex generation efficiency $\eta_{LR}$ as a function of the vacuum wavelength $\lambda = 2\pi c/\omega$ and the beam width $w$. The upper triangular gray region has been omitted since it corresponds to evanescent Bessel beams where $w < \lambda$. The nonparaxial regime where $w \simeq \lambda$ corresponds to the region close to the black line limiting the gray evanescent region and here $\eta_{LR}$ is sensibly different from zero at least for wavelengths smaller than $\lambda _0$. This is the above discussed vortex generation for nonparaxial fields occurring in dielectric slabs (since ${\rm Re} \: \varepsilon > 0$ for $\lambda > \lambda_0$, see Fig.4(a)). In the paraxial regime $w > \lambda$, $\eta_{LR}$ is very small except for a region surrounding the wavelength $\lambda_0$ where $\eta_{LR}$ fades as $w$ increases. This shows that vortex generation is effectively operated by the considered ITO slab in the paraxial regime only for wavelength close to the ENZ one. In Fig.4(c) we plot $\eta_{LR}$ at $\lambda_0$ as a function of the slab thickness $L$ and the beam width $w$. Note that the vortex generation efficiency in this example is rather small as a consequence of the large imaginary part of $\varepsilon$ at $\lambda_0$ which prevents the full ENZ condition $|\varepsilon| \ll 1$ to be fulfilled. In addition, the large slab losses restrict vortex generation to very thin slabs $L \lesssim 0.3 \: \mu {\rm m} \simeq \frac{1}{4} \lambda_0 $.

In the second example we consider a sodium (Na) slab of thickness $L=100 \: {\rm nm}$ whose Drude parameters are $\epsilon_\infty = 1$, $\omega_p = 8.2 \times 10^{15}$ Hz and $\gamma = 0.003 \: \omega_p$ \cite{Forstm}, for which $\lambda_0 = 222 \: {\rm nm}$ and ${\rm Im} \left[ \varepsilon (\lambda_0) \right] = 0.003$. In Figs.5(a) and 5(b) we focus on the ultraviolet spectral band $200 \: {\rm nm} < \lambda < 245 \: {\rm nm}$ and, in analogy with Fig.4, we plot the permittivity and vortex generation efficiency of the considered Na slab. This example shows all the above discussed features of slab vortex generation but its efficiency is larger than in the first example. In Fig.5(c) we plot $\eta_{LR}$ at $\lambda_0$ as a function of the slab thickness $L$ and the beam width $w$ and, as compared with Fig.4(c), it reveals both the larger vortex generation efficiency and the persistence of the phenomenon up to very wide paraxial fields (where $w \simeq 18 \lambda_0$ for $w=4000 \: {\rm nm}$). This is due to the small imaginary part of $\varepsilon$ at $\lambda_0$ which allows the Na slab to host the genuine ENZ regime $|\varepsilon| \ll 1$ close to $\lambda_0$. The large impact of losses in the ENZ restrict vortex generation to sub-wavelength slabs of thickness $L \lesssim 100 \: {\rm nm} \simeq \frac{1}{2} \lambda_0 $.

In conclusion we have shown that a sub-wavelength thick slab supports efficient vortex generation in nonparaxial fields. If the slab is in the ENZ regime, its ability to convert spin into angular orbital momentum extends to the paraxial regime. This is a remarkable effect in view of the homogeneity of the slab and its extremely small thickness which is made possible by the conversion of a paraxial field into a nonparaxial one operated by the ENZ slab. As compared to standard vortex generation techniques based on metasurfaces, our method is based on a very simple setup, the homogenous ENZ slab, which does not require micro-fabrication and which can be exploited even for very small ultraviolet wavelength where metamaterial are unavailable and metals have their plasma frequency. As a consequence, our method can be the platform for a novel generation of devices for manipulating the radiation angular momentum.

A.C. and C.R. acknowledge support from U.S. Army International Technology Center Atlantic for financial support (Grant No. W911NF-14-1-0315).


\begin{thebibliography} {aa}
\bibitem{Bliok1} K. Y. Bliokh, F. J. Rodriguez-Fortuno, F. Nori, and A. V. Zayats, Nat. Photon. \textbf{9}, 796-808 (2015).
\bibitem{Zhaooo} Y. Zhao, J. S. Edgar, G. D. M. Jeffries, D. McGloin, and D. T. Chiu, Phys. Rev. Lett. \textbf{99}, 073901 (2007).
\bibitem{Bliok2} K. Y. Bliokh, E. A. Ostrovskaya, M. A. Alonso, O. G. Rodríguez-Herrera, D. Lara, and C. Dainty, Opt. Express \textbf{19}, 26132-26149 (2011).
\bibitem{Ciatt1} A. Ciattoni, G. Cincotti, and C. Palma, J. Opt. Soc. Am A \textbf{20}, 163-171 (2003).
\bibitem{Brass1} E. Brasselet, Y. Izdebskaya, V. Shvedov, A. S. Desyatnikov, W. Krolikowski, and Y. S. Kivshar, Opt. Lett. \textbf{34}, 1021-1023 (2009).
\bibitem{Mannii} F. Manni, K. G. Lagoudakis, T. K. Para\"{\i}so, R. Cerna, Y. L\'{e}ger, T. C. H. Liew, I. A. Shelykh, A. V. Kavokin, F. Morier-Genoud, and B. Deveaud-Pl\'{e}dran, Phys. Rev. B \textbf{83}, 241307(R) (2011).
\bibitem{Khilo1} N. A. Khilo, E. S. Petrova, and A. A. Ryzhevich, Quantum Elect. \textbf{31}, 85-89 (2001).
\bibitem{Yavors} M. Yavorsky, and E. Brasselet, Opt. Lett. \textbf{37}, 3810-3812 (2012).
\bibitem{Shitri} N. Shitrit, I. Yulevich, E. Maguid, D. Ozeri, D. Veksler, V. Kleiner, and E. Hasman, Science \textbf{340}, 724-726 (2013).
\bibitem{Liiiii} G. Li, M. Kang, S. Chen, S. Zhang, E. Y.-B. Pun, K. W. Cheah, and J. Li, Nano Lett. \textbf{13}, 4148-4151 (2013).
\bibitem{Hakoby} D. Hakobyan, H. Magallanes, G. Seniutinas, S. Juodkazis, and E. Brasselet, Adv. Optical Mater. \textbf{4}, 306–312 (2016).
\bibitem{Yanggg} Y. Yang, W. Wang, P. Moitra, I. I. Kravchenko, D. P. Briggs, and J. Valentine, Nano Lett. \textbf{14} 1394-1399 (2014).
\bibitem{Chennn} S. Chen, Y. Cai, G. Li, S. Zhang, and K. W. Cheah, Laser Photon. Rev. \textbf{10}, 322–326 (2016).
\bibitem{Silve1} M. Silveirinha, and N. Engheta, Phys. Rev. Lett. \textbf{97}, 157403 (2006).
\bibitem{Aluuu1} A. Al\'u, M. Silveirinha, A. Salandrino, and N. Engheta, Phys. Rev. B \textbf{75}, 155410 (2007).
\bibitem{Ciatt2} A. Ciattoni, C. Rizza, and E. Palange, Phys. Rev. A \textbf{81}, 043839 (2010).
\bibitem{Argyr1} C. Argyropoulos, P. Y. Chen, G. D’Aguanno, N. Engheta, and A. Al\'u,  Phys. Rev. B \textbf{85}, 045129 (2012).
\bibitem{Ciatt3} A. Ciattoni, C. Rizza, A. Marini, A. Di Falco, D. Faccio, and M. Scalora, Laser Photon. Rev. \textbf{10}, 517–525 (2016).
\bibitem{Alammm} M. Z. Alam, I. De Leon, and R. W. Boyd, Science \textbf{352}, 795-797 (2016).
\bibitem{Ciatt4} A. Ciattoni,  A. Marini, and C. Rizza, Opt. Lett. \textbf{41}, 3102-3105 (2016).
\bibitem{Zhuuuu} W. Zhu, and W. She, Opt. Lett. \textbf{40}, 2961-2964 (2015).
\bibitem{Forstm} F. Forstmann, and R. R. Gerhardt, {\it Metal Optics Near the Plasma Frequency}, Springer-Verlag, 1986.

\end{thebibliography}
\end{document}